\documentclass[manuscript]{aastex}
\usepackage{rotating,graphicx}

\shorttitle{Interplay between Scattering and Adiabatic Focusing}

\shortauthors{He \& Wan}

\begin{document}

\title{Propagation of Solar Energetic Particles in the outer Heliosphere: Interplay between Scattering and Adiabatic Focusing \\}

\author{H.-Q. He\altaffilmark{1,2,3} and W. Wan\altaffilmark{1,2,3}}

\altaffiltext{1}{Key Laboratory of Earth and Planetary Physics,
Institute of Geology and Geophysics, Chinese Academy of Sciences,
Beijing 100029, China; hqhe@mail.iggcas.ac.cn}

\altaffiltext{2}{Innovation Academy for Earth Science, Chinese
Academy of Sciences, Beijing 100029, China}

\altaffiltext{3}{Beijing National Observatory of Space Environment,
Institute of Geology and Geophysics, Chinese Academy of Sciences,
Beijing 100029, China}

\begin{abstract}
The turbulence and spatial nonuniformity of the guide magnetic field
cause two competitive effects, namely, the scattering effect and the
adiabatic focusing effect, respectively. In this work, we
numerically solve the five-dimensional Fokker-Planck transport
equation to investigate the radial evolutions of these important
effects undergone by the solar energetic particles (SEPs)
propagating through interplanetary space. We analyze the interplay
process between the scattering and adiabatic focusing effects in the
context of three-dimensional propagation, with special attention to
the scenario of the outer heliosphere, in which some peculiar SEP
phenomena are found and explained. We also discuss the radial
dependence of the SEP peak intensities from the inner through outer
heliosphere, and conclude that it cannot be simply described by a
single functional form such as $R^{-\alpha}$ ($R$ is radial
distance), which is often used.
\end{abstract}

\keywords{diffusion -- turbulence -- interplanetary medium -- Sun:
particle emission -- Sun: magnetic fields -- Sun: heliosphere}

\clearpage

\section{Introduction}
Solar energetic particles (SEPs) are produced near the Sun during
solar eruptive events and will transport in the interplanetary space
filled with turbulent magnetic fields. The SEP events observed in
the heliosphere provide fundamental information regarding particle
acceleration and transport mechanisms, which are enigmatic problems
of long-standing importance in space physics, plasma physics, and
astrophysics. Therefore, the SEPs can serve as convenient messengers
for us to achieve a better understanding of the fundamental
behaviors of energetic particles in extreme astrophysical
environments including dynamical plasma media and waves and
turbulent magnetic fields. In the upcoming era of Parker Solar Probe
and Solar Orbiter, it is expectable to make significant advances in
understanding the physical mechanisms of particle origin,
acceleration, and transport.

The charged energetic particles in magnetic turbulence experience
scattering and diffusion processes both parallel and perpendicular
to the large-scale guide magnetic field. Parallel diffusion of
charged particles along the mean magnetic field has been extensively
investigated
\citep[e.g.,][]{Droge2000,Shalchi2005,Shalchi2006,He2012a}.
Recently, perpendicular diffusion of charged particles across the
guide magnetic field has also been intensely studied
\citep[e.g.,][]{Zhang2009,He2011,He2015,Droge2014,Shalchi2010,Shalchi2019}.
The parallel mean free path $\lambda_{\parallel,0}$ of charged
particles in a uniform mean magnetic field can be written as
\citep{Jokipii1966,Hasselmann1968,Earl1974}
\begin{equation}
\lambda_{\parallel,0}=\frac{3v}{8}\int_{-1}^{+1}\frac{(1-\mu^{2})^{2}}{D_{\mu\mu}}d\mu,
\label{parallel-path}
\end{equation}
where $D_{\mu\mu}$ is the pitch-angle diffusion coefficient.
However, most of the astrophysical magnetic fields including the
interplanetary magnetic fields are often nonuniform. This spatially
varying mean magnetic field gives rise to the adiabatic focusing
effect of charged energetic particles
\citep{Roelof1969,Earl1976,Bieber1990,Ruffolo1995,Bieber2002,Schlickeiser2008}.
The presence of the adiabatic focusing effect causes coherent
spatial particle streaming along the large-scale guide magnetic
field in magnetostatic turbulence. In the scenario of the inner
heliosphere, the adiabatic focusing effect is very important and
must be taken into account when we analyze the diffusion and
transport processes of SEPs in the interplanetary space
\citep{He2012a}. In general, the adiabatic focusing length $L(z)$
can be defined via
\begin{equation}
\frac{1}{L(z)}=-\frac{\partial \ln B(z)}{\partial
z}=-\frac{1}{B(z)}\frac{\partial B(z)}{\partial z},\label{focussing}
\end{equation}
where $B(z)$ is the mean magnetic field with direction $z$. As we
can see, the focusing length $L(z)$ is positive in a diverging guide
field and is negative in a converging guide field.

In previous studies, most authors theoretically investigated the
effects of magnetic adiabatic focusing on the parallel diffusion
coefficients of charged particles
\citep{Beeck1986,Bieber1990,Ruffolo1995,Kota2000,Schlickeiser2008,Litvinenko2012,Shalchi2013,He2014}.
They usually focused on calculating the modifications of spatial
diffusion coefficients of charged particles transporting in
nonuniform guide magnetic fields. During the derivations, a number
of useful and universal approximation and perturbation methods have
been presented in the literature. However, it is quite scarce to see
the investigations within the community that directly analyze the
effects of adiabatic focusing on the transport and distribution of
SEPs in the inner and outer heliosphere by presenting the
time-intensity profiles of radial evolutions of SEPs, especially in
the physical scenario of three-dimensional propagation including
perpendicular diffusion. Undoubtedly, such investigation tasks are
very important for us to achieve a clear and explicit understanding
of the effects of adiabatic focusing on the SEP diffusion,
transport, and distribution. In addition, performing such tasks can
bring us a detailed illustration of the interplay process between
the adiabatic focusing effect and the particle scattering effect. To
this aim, it is necessary to numerically simulate the
three-dimensional transport processes of SEPs in the inner and outer
heliosphere, since the multidimensional focused transport equation
is very difficult to be solved analytically.

In this work, we numerically solve the five-dimensional
Fokker-Planck transport equation that incorporates all of the
essential transport mechanisms including perpendicular diffusion. We
analyze the simulation results and investigate the radial evolutions
of SEP time-flux profiles from the inner through outer heliosphere.
The effects of adiabatic focusing and SEP scattering and the
interplay process between them in the interplanetary magnetic fields
will be discussed. Some peculiar SEP phenomena including SEP
``floods" (previously ``reservoirs") found in the inner and outer
heliosphere will also be analyzed and discussed. This paper is
structured as follows. In Section 2, we present the numerical model
used in this work, i.e., the five-dimensional Fokker-Planck focused
transport equation, and the relevant simulation method for
numerically solving the equation. In Section 3, we present the
simulation results and discuss the physical mechanisms, with special
attention to the effects of adiabatic focusing and particle
scattering. We also present multi-spacecraft observations for
comparison with simulations. In Section 4, a summary of our results
will be provided.

\section{Numerical Model and Method}
The five-dimensional time-dependent Fokker-Planck transport equation
for the gyrophase-averaged SEP distribution function
$f(\textbf{x},\mu,p,t)$, which incorporates the effects of adiabatic
focusing and particle scattering in pitch-angle cosine $\mu$, can be
written as
\citep[e.g.,][]{Schlickeiser2002,Zhang2009,He2011,He2015,He2017b,Droge2014,Droge2016}
\begin{eqnarray}
{}&&\frac{\partial f}{\partial t}+\mu v\frac{\partial f}{\partial
z}+{\bf V}^{sw}\cdot\nabla f+\frac{dp}{dt}\frac{\partial f}{\partial
p}+\frac{d\mu}{dt}\frac{\partial f}{\partial \mu}  \nonumber\\
{}&&-\frac{\partial}{\partial\mu}\left(D_{\mu\mu}\frac{\partial
f}{\partial \mu}\right)-\frac{\partial}{\partial
x}\bigg(\kappa_{xx}\frac{\partial f}{\partial x}\bigg)
-\frac{\partial}{\partial y}\left(\kappa_{yy}\frac{\partial
f}{\partial y}\right)=Q({\bf x},p,t).  \label{transport-equation}
\end{eqnarray}
In the above Fokker-Planck transport equation, $\textbf{x}$ denotes
the spatial position of particles, $z$ denotes the spatial
coordinate along the guide magnetic field line $B(z)$, $p$ is
particle's momentum, $t$ is time, $v$ is particle's velocity,
$\textbf{V}^{sw}$ is solar wind speed, $\kappa_{xx}$ and
$\kappa_{yy}$ denote perpendicular diffusion coefficients, and
$Q(\textbf{x},p,t)$ denotes particle source term. The term $dp/dt$,
describing the adiabatic cooling effect, can be written as
\begin{equation}
\frac{dp}{dt}=-p\left[\frac{1-\mu^2}{2}\left(\frac{\partial
V^{sw}_x}{\partial x}+\frac{\partial V^{sw}_y}{\partial
y}\right)+\mu^2\frac{\partial V^{sw}_z}{\partial z}\right].
\label{adiabatic-cooling}
\end{equation}
The term $d\mu/dt$, representing the effect of magnetic adiabatic
focusing and the divergence of solar wind flows, can be written as
\begin{eqnarray}
\frac{d\mu}{dt}&=&\frac{1-\mu^2}{2}\left[-\frac{v}{B}\frac{\partial
B}{\partial z}+\mu \left(\frac{\partial V^{sw}_x}{\partial
x}+\frac{\partial V^{sw}_y}{\partial y}-2\frac{\partial
V^{sw}_z}{\partial z}\right)\right]     \nonumber\\
{}&=&\frac{1-\mu^2}{2}\left[\frac{v}{L}+\mu\left(\frac{\partial
V^{sw}_x}{\partial x}+\frac{\partial V^{sw}_y}{\partial
y}-2\frac{\partial V^{sw}_z}{\partial z}\right)\right],
\label{magnetic-focusing}
\end{eqnarray}
where $B$ denotes the guide interplanetary magnetic field, and $L$
denotes the magnetic focusing length.

Accordingly, the radial mean free path $\lambda_{r}$ can be
expressed as
\begin{equation}
\lambda_{r}=\lambda_{\parallel}\cos^{2}\psi. \label{radial-path}
\end{equation}
Here, $\psi$ denotes the angle between the local magnetic field
direction and the radial direction. We utilize a pitch-angle
diffusion coefficient with the following form
\citep[e.g.,][]{Beeck1986,Zhang2009,He2011}
\begin{equation}
D_{\mu\mu}^{r}=D_{\mu\mu}/\cos^{2}\psi=D_{0}vR_{d}^{-1/3}\left(|\mu|^{q-1}+h\right)(1-\mu^{2}),
\label{diffusion-coefficient}
\end{equation}
where $D_{0}$ denotes the magnetic turbulence strength, $R_{d}$
denotes the particle rigidity, $h$ is a parameter set to describe
the particle scattering ability through $90^{\circ}$ pitch-angle,
and $q$ is a parameter relevant to the power spectrum of the
magnetic turbulence in inertial range, which is set to be $5/3$ in
this work. Recently, a more accurate expression for the pitch-angle
scattering coefficient was derived systematically from non-linear
diffusion theory which also provides a non-vanishing scattering
coefficient at $\mu=0$ \citep{Shalchi2009}. In addition, note that
the focusing effect can alter the pitch-angle scattering effects
itself \citep{Tautz2014}.

We employ the so-called time-backward Markov stochastic process
approach to numerically solve the five-dimensional Fokker-Planck
transport Equation (\ref{transport-equation}). Through this
approach, the Fokker-Planck Equation (\ref{transport-equation}) can
be readily transformed into five time-backward stochastic
differential equations (SDEs) as in the following:
\begin{eqnarray}
dX &=& \sqrt{2\kappa_{xx}}dW_{x}(s)-V_{x}^{sw}ds  \nonumber\\
dY &=& \sqrt{2\kappa_{yy}}dW_{y}(s)-V_{y}^{sw}ds  \nonumber\\
dZ &=& -(\mu V+V_{z}^{sw})ds  \nonumber\\
d\mu &=& \sqrt{2D_{\mu\mu}}dW_{\mu}(s)  \nonumber\\
{}&& -\frac{1-\mu^{2}}{2}\left[\frac{V}{L}+\mu\left(\frac{\partial
V_{x}^{sw}}{\partial x}+\frac{\partial V_{y}^{sw}}{\partial
y}-2\frac{\partial V_{z}^{sw}}{\partial z}\right)\right]ds  \nonumber\\
{}&& +\left(\frac{\partial D_{\mu\mu}}{\partial
\mu}+\frac{2D_{\mu\mu}}{M+\mu}\right)ds  \nonumber\\
dP &=& P\left[\frac{1-\mu^{2}}{2}\left(\frac{\partial
V_{x}^{sw}}{\partial x}+\frac{\partial V_{y}^{sw}}{\partial
y}\right)+\mu^{2}\frac{\partial V_{z}^{sw}}{\partial z}\right]ds,
\label{eq:stochastic-process}
\end{eqnarray}
where $(X,Y,Z)$ denotes the particle pseudo-position, $V$ denotes
the particle pseudo-speed, $P$ denotes the particle pseudo-momentum,
and $W_{x}(t)$, $W_{y}(t)$, and $W_{\mu}(t)$ denote the Wiener
processes. The quantity of the gyrophase-averaged particle
distribution function $f(\textbf{x},\mu,p,t)$ can be numerically
obtained from the five SDEs (\ref{eq:stochastic-process}). In the
numerical simulations, we trace a number of particles back to the
initial time of the physical system. In the statistical analyses, we
only take into account those ``effective" particles which arrive at
the source region at the initial time.

The particle source term $Q(\textbf{x},p,t)$ in the Fokker-Planck
transport Equation (\ref{transport-equation}), which serves as an
inner injection boundary of particles in the simulations, is assumed
to be as \citep{Reid1964}
\begin{equation}
Q(R\leqslant0.05AU,\theta,\phi,E_{k},t)=\frac{C}{t}\frac{E_{k}^{-\gamma}}{p^{2}}
\exp\left(-\frac{\tau_{c}}{t}-\frac{t}{\tau_{L}}\right)\xi(\theta,\phi),
\label{source}
\end{equation}
where $\gamma$ denotes the spectral index of source region particles
which is chosen to be $3$, $\tau_{c}$ and $\tau_{L}$ denote the time
quantities that determine the particle injection profile in source
regions, and $\xi(\theta,\phi)$ is a function controlling the
spatial variation (longitude and latitude) of particle injection
strength in source regions. We note that the SEP injection model
shown in Equation (\ref{source}) can be used to describe either the
SEP release from solar flares or the SEP injections from shocks
driven by coronal mass ejections (CMEs) in the corona. This SEP
source model is particularly suitable for describing the short-lived
injections of high-energy particles released near the Sun. In this
work, we concentrate on the SEP time-intensity profiles in the
prompt component of SEP events.

In addition, we set an outer boundary at radial distance $R=50$ AU
for absorbing the particles when they hit the boundary. For the
interplanetary conditions, we typically use a constant solar wind
speed of $V^{sw}=400~km~s^{-1}$, and a spiral-type interplanetary
magnetic field with strength $B=5 nT$ at $1 AU$. For each SEP case,
we simulate $3\times10^{7}$ test particles on a super-computer
cluster. During the data analyses of the simulation results, we
adopt an arbitrary unit for presenting the time-flux profiles of
particles instead of using the usual
$cm^{-2}-s^{-1}-sr^{-1}-MeV^{-1}$, because of the consideration of
convenience in plotting figures.

\section{Numerical Results and Discussion}
\begin{figure}
 \epsscale{1.0}
 \plotone{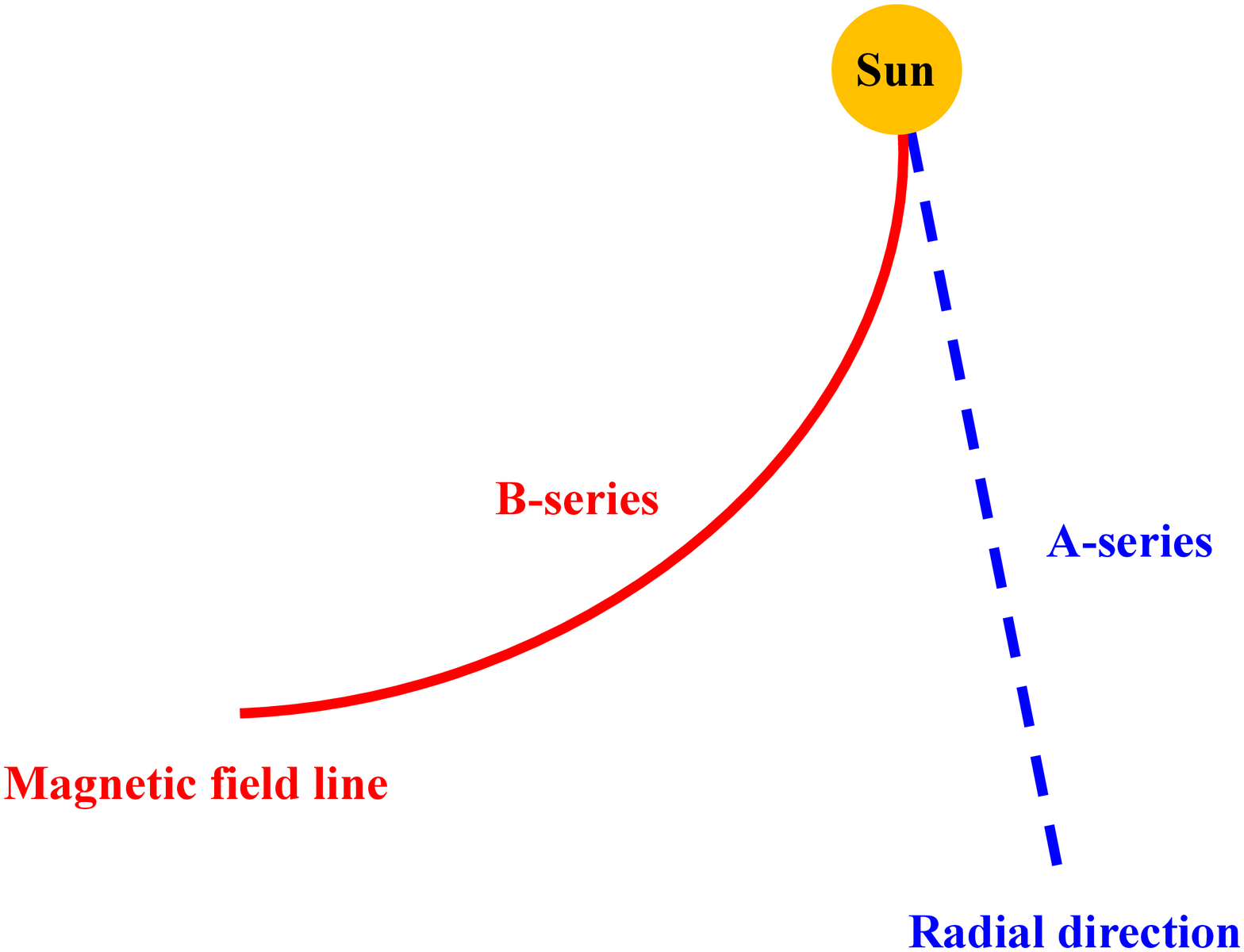}
 \caption{Illustrative sketch to show the alignment scenarios of the
 spacecraft fleet in the heliosphere. The blue dashed line denotes the
 radial direction along which the A-series spacecraft are aligned with
 different radial distances. The red solid curve indicates the interplanetary
 magnetic field line along which the B-series spacecraft are aligned with different
radial distances. The heliocentric radial distances of the
spacecraft fleet in each alignment are in sequence: $0.25$, $0.4$,
$0.6$, $0.8$, $1.0$, $1.5$, $2.0$, $2.5$, $3.0$, $3.5$, $4.0$,
$4.5$, and $5.0$ AU. Both the spacecraft fleet and the SEP sources
are located at $90^{\circ}$ colatitude. \label{scenarios}}
\end{figure}

We first present an illustrative sketch, i.e., Figure
\ref{scenarios}, to show the physical scenarios discussed in this
paper. The blue dashed line in Figure \ref{scenarios} denotes the
radial direction along which the A-series spacecraft are aligned
with different radial distances. The red solid curve indicates the
interplanetary magnetic field line along which the B-series
spacecraft are aligned with different radial distances. Both the
radial direction line and the magnetic field line originate from the
same SEP source near the Sun. The heliocentric radial distances of
the spacecraft fleet in each alignment scenario (A-series and
B-series) are in sequence: $0.25$, $0.4$, $0.6$, $0.8$, $1.0$,
$1.5$, $2.0$, $2.5$, $3.0$, $3.5$, $4.0$, $4.5$, and $5.0$ AU. In
addition, both the spacecraft fleet and the SEP sources are located
at $90^{\circ}$ colatitude. In the simulations, the SEP sources are
set to be with limited coverages, i.e., $45^{\circ}$ or $70^{\circ}$
in longitude and latitude.

\begin{figure}
 \epsscale{0.7}
 \plotone{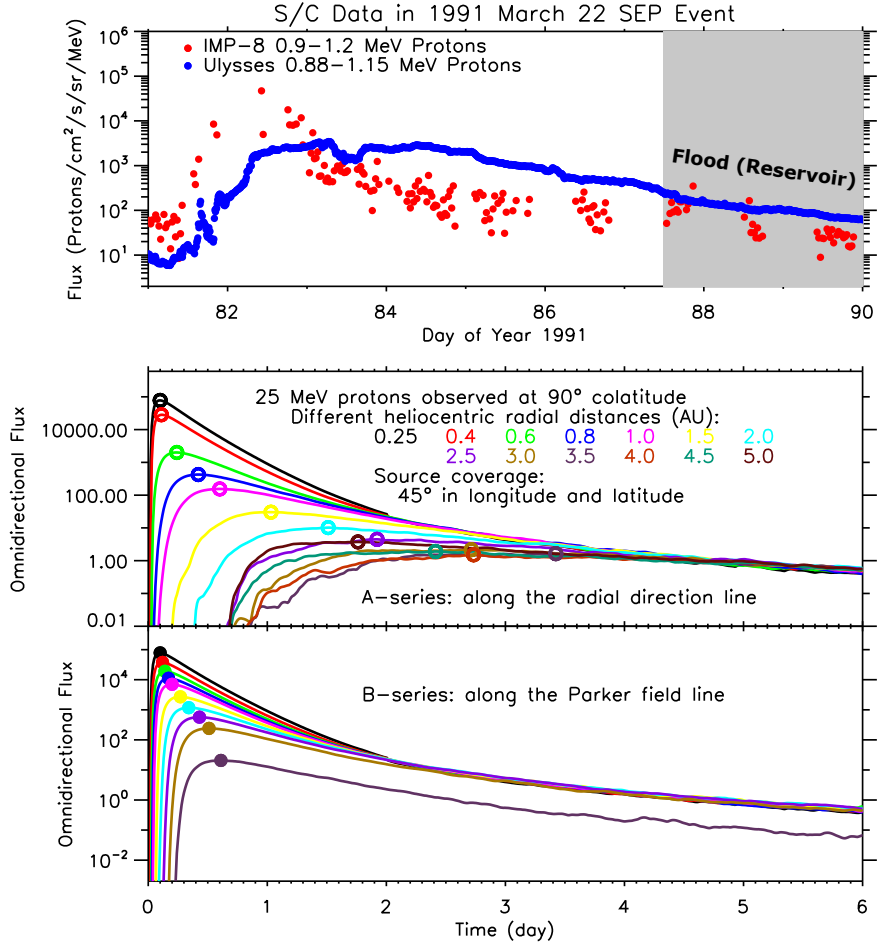}
% \vspace{0.5cm}
 \caption{Observations (top) and simulation results (middle and bottom) of the radial evolutions
 of SEP time-flux profiles from the inner through outer heliosphere. Top panel: 0.9-1.2 MeV protons
 (red circles, 30-min average) observed by IMP-8 at 1.0 AU and 0.88-1.15 MeV protons (blue circles, 10-min average)
 observed by Ulysses at 2.5 AU during the 1991 March 22 (day of year 81) SEP event. For the simulations, the middle
 panel denotes the SEP scenario along the radial direction, and the bottom panel denotes the scenario
 along the interplanetary magnetic field line originating from the SEP source. The particles are $25$ MeV protons,
 and the SEP source is $45^{\circ}$ wide in latitude and longitude. The different colors of the time-flux profiles
 indicate the simulations at different radial distances: $0.25$, $0.4$, $0.6$, $0.8$, $1.0$, $1.5$, $2.0$, $2.5$,
 $3.0$, $3.5$, $4.0$, $4.5$, and $5.0$ AU. The open and solid circles on the time-flux profiles in
 the middle (A-series) and bottom (B-series) panels, respectively, denote the peak fluxes of the corresponding SEP cases. \label{25MeV}}
\end{figure}

Figure \ref{25MeV} presents the observations (top panel) and
simulation results (middle and bottom panels) of the radial
evolutions of SEP time-flux profiles from the inner through outer
heliosphere. In the top panel, the red solid circles denote the
time-flux profiles of 0.9-1.2 MeV protons (30-min average) observed
by IMP-8 spacecraft at 1.0 AU, and the blue solid circles denote the
time-flux profiles of 0.88-1.15 MeV protons (10-min average)
observed by Ulysses spacecraft at 2.5 AU. The measurements of
particle fluxes on both spacecraft were made during the 1991 March
22 (day of year 81) SEP event. Note that there are data gaps in the
time-flux profiles. As we can see, the phase of the rise and the
peak of the particle fluxes is quite different at IMP-8 and Ulysses,
but during the late phase (indicated by gray-shaded area), the
fluxes at 1.0 AU and at 2.5 AU are very nearly equal and evolve
similarly in time with almost the same decay rates. This particle
behavior is the so-called SEP ``flood" (previously ``reservoir")
phenomenon \citep{McKibben1972,Roelof1992,He2017a}. Note that the
SEP ``floods" (previously ``reservoirs") are observed in both low
and high energy particle data, and also in both proton data and
electron and heavy-ion data. This SEP phenomenon is detected by
spacecraft at different heliolongitudes, heliolatitudes, and radial
distances. The middle and bottom panels of Figure \ref{25MeV}
present the simulation results of two different SEP scenarios. In
the middle panel, the time-flux profiles are observed along the
radial direction. In the bottom panel, the time-flux profiles are
detected along the Parker-type interplanetary magnetic field line.
In the simulation scenarios of middle and bottom panels, the
coverage of particle source is set to be $45^{\circ}$ both in
longitude and latitude. In these two panels, the different colors of
the time-flux profiles indicate the simulation results obtained at
different radial distances: $0.25$, $0.4$, $0.6$, $0.8$, $1.0$,
$1.5$, $2.0$, $2.5$, $3.0$, $3.5$, $4.0$, $4.5$, and $5.0$ AU. For
both scenarios of spacecraft alignment (``A-series" and
``B-series"), the SEP diffusion coefficients are set as follows: the
radial mean free path $\lambda_{r}=0.25$ AU (corresponding to the
parallel mean free path $\lambda_{\parallel}=0.5$ AU at 1 AU), and
the perpendicular mean free paths $\lambda_{x}=\lambda_{y}=0.006$
AU. Note that the values of the parallel and perpendicular mean free
paths are based on the recent results of observations and theories
regarding the parallel and perpendicular diffusion coefficients of
energetic charged particles in the interplanetary space
\citep[e.g.,][]{Bieber1994,Droge2000,Matthaeus2003,Bieber2004,He2012a,He2012b}.
In the middle panel, we can see that the SEP intensities
monotonically decrease with increasing radial distances in the inner
heliosphere and as far as $\sim4.0$ AU, beyond where the particle
intensities counterintuitively and gradually increase with the
increasing radial distances up to at least $5.0$ AU. The reason is
that at relatively large radial distances, e.g., at $\gtrsim3.2$ AU
computed assuming a solar wind speed of $400~km~s^{-1}$, the larger
the radial distance of the observer is, the closer the magnetic
footpoint of the observer is to the SEP source, and consequently in
the sense of longitudinal distance, the higher the particle flux
observed will be. As one can see, during the late phases, the
particle intensities of the SEP events present nearly equal values
and evolve similarly in time with almost the same decay rates. This
evolution feature of SEP events is the so-called SEP ``flood"
(previously ``reservoir") phenomenon as described in the top panel.
Therefore, we successfully reproduce this famous SEP phenomenon by
simulating the SEP three-dimensional transport process from the
inner through outer heliosphere. In the bottom panel of Figure
\ref{25MeV}, the particle intensities gradually decrease with
increasing radial distances in the inner heliosphere and up to
$\sim3.0$ AU. However, at $\gtrsim3.5$ AU, the SEP intensity
abruptly decreases to a quite low value. Furthermore, the particles
almost ``disappear" at radial distances $\gtrsim4.0$ AU. The reason
is that at large radial distances from the Sun, the adiabatic
focusing effect is significantly reduced \citep{He2012a}, and as a
result, the particles considerably deviate from the primary magnetic
field lines, which they previously followed during the early stage.
As one can see in the bottom panel of Figure \ref{25MeV}, the
so-called SEP ``flood" (previously ``reservoir") phenomenon
\citep{McKibben1972,Roelof1992,He2017a} is successfully reproduced
in our three-dimensional transport modeling of SEP propagation from
the inner to the outer heliosphere. We note that the open and solid
circles on the time-flux profiles in the middle (A-series) and
bottom (B-series) panels of Figure \ref{25MeV}, respectively, denote
the peak fluxes of the corresponding SEP cases. These peak fluxes
and their radial evolutions will be discussed later.

\begin{figure}
 \epsscale{1.0}
 \plotone{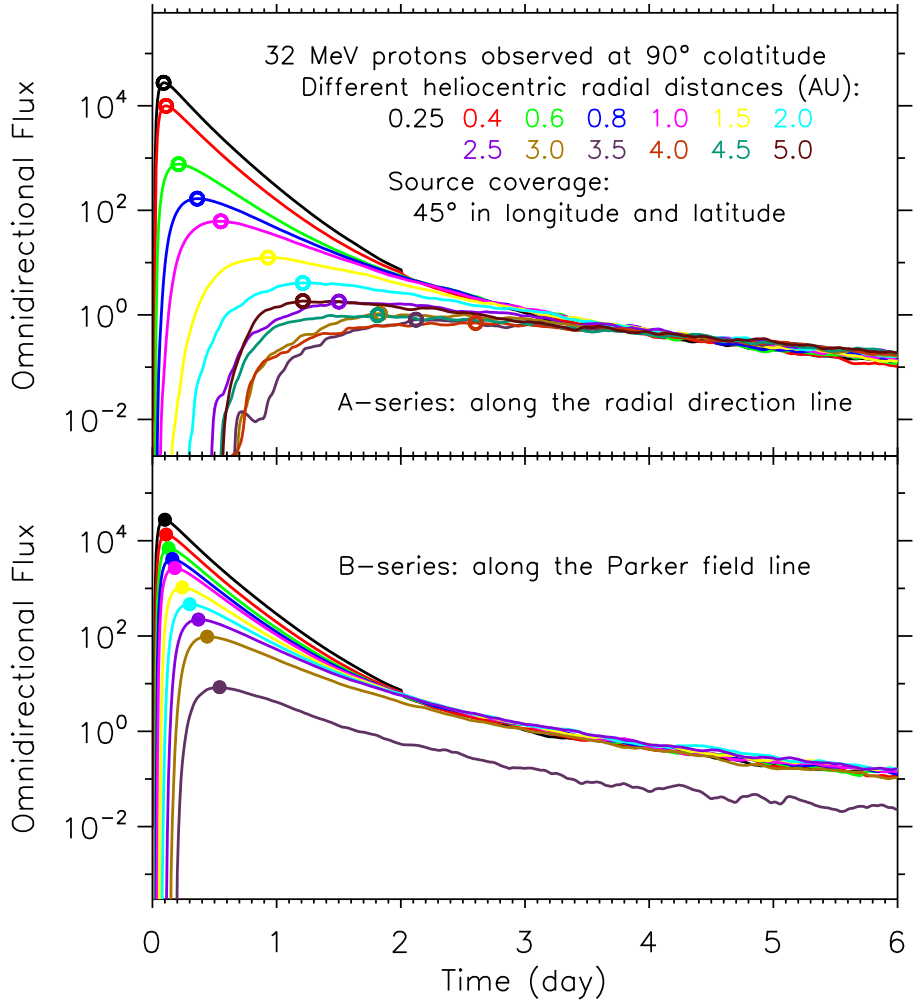}
 \vspace{0.5cm}
 \caption{Same as the simulations (middle and bottom panels) in Figure \ref{25MeV} except for $32$ MeV protons. \label{32MeV}}
\end{figure}

Figure \ref{32MeV} shows the numerical simulation results of the
radial variations of the time-flux profiles of $32$ MeV solar
protons transporting from the inner to the outer heliosphere.
According to the recent results of observations and theories
\citep[e.g.,][]{Droge2000,Matthaeus2003,Bieber2004,He2012a,He2012b},
the diffusion coefficients of the energetic particles are typically
set as follows: the radial mean free path $\lambda_{r}=0.28$ AU
(corresponding to the parallel mean free path
$\lambda_{\parallel}=0.56$ AU at 1 AU), and the perpendicular mean
free paths $\lambda_{x}=\lambda_{y}=0.007$ AU. Other physical
conditions and modeling parameters are set the same as the
simulations in the middle and bottom panels of Figure \ref{25MeV}.
In the upper panel of Figure \ref{32MeV}, one can see that the SEP
fluxes monotonically decline with increasing radial distances in the
inner heliosphere and up to $\sim4.0$ AU. However, afterwards the
SEP intensities gradually increase with the increasing radial
distances as far as at least $5.0$ AU. This peculiar SEP phenomenon
results from the fact that at relatively large radial distances,
e.g., at $\gtrsim3.2$ AU in this work, the larger the observer's
radial distance is, the closer the observer's magnetic footpoint is
to the SEP source in longitude, and as a result, the higher the SEP
intensity measured will be. In the lower panel of Figure
\ref{32MeV}, the SEP fluxes gradually decline with increasing radial
distances in the inner heliosphere and up to $\sim3.0$ AU.
Nevertheless, the SEP flux dramatically declines to a very low
magnitude at $\gtrsim3.5$ AU. Further, the SEP intensities almost
``vanish away" at radial distances $\gtrsim4.0$ AU. The reason is
that at large radial distances, the effect of adiabatic focusing is
largely reduced \citep{He2012a}, and consequently the SEPs
significantly deviate from the primary magnetic field lines
originating from the limited source region. Note that in both
panels, the SEP ``flood" (previously ``reservoir") phenomenon is
reproduced. The open and solid circles on the time-flux profiles in
the upper and lower panels of Figure \ref{32MeV}, respectively,
indicate the peak intensities of the corresponding SEP cases.

\begin{figure}
 \epsscale{1.0}
 \plotone{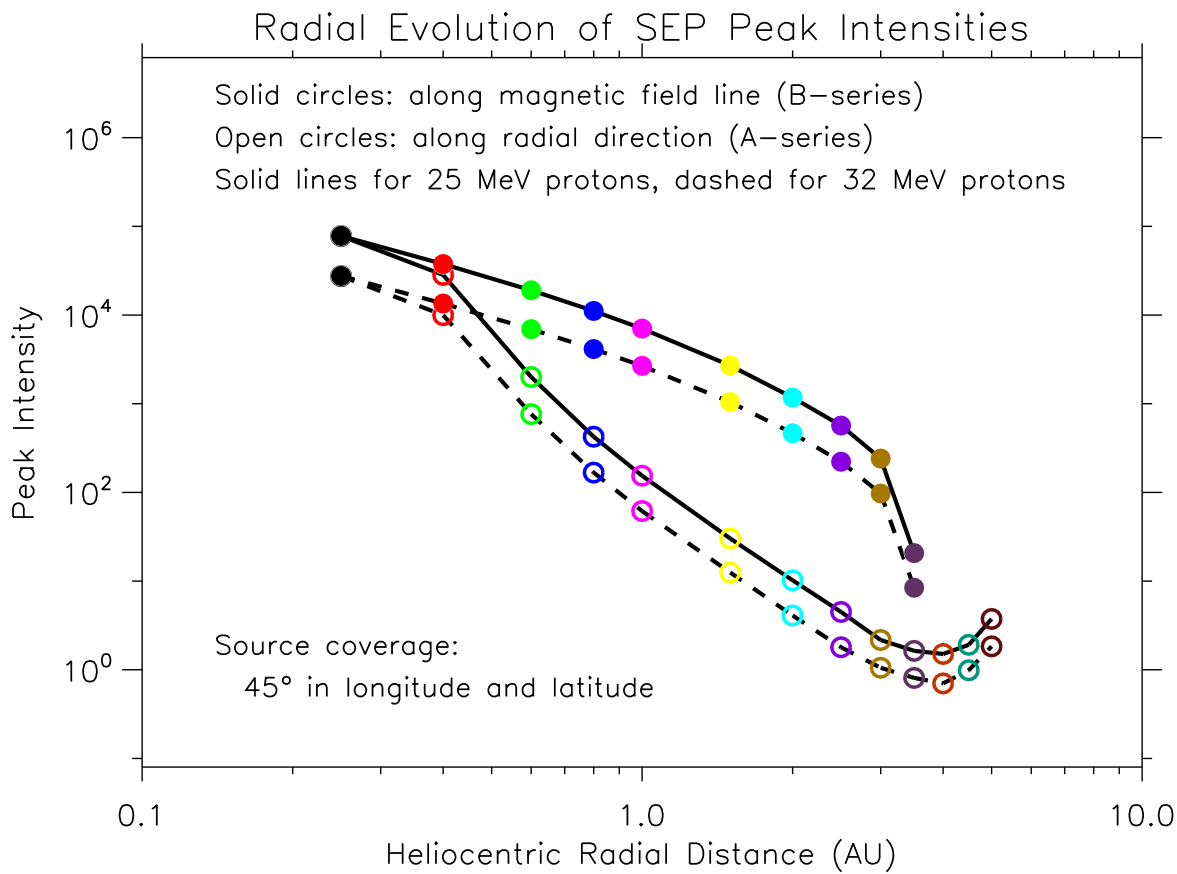}
 \caption{Radial evolutions of the peak fluxes of 25 MeV (solid lines) and 32 MeV (dashed lines) proton events, extracted
 from the simulation results in Figures \ref{25MeV} and \ref{32MeV}, respectively. The solid and open circles denote the
 SEP peak fluxes observed along the magnetic field line (B-series) and along the radial direction (A-series),
 respectively. The SEP source is $45^{\circ}$ wide in latitude and longitude. The radial evolution of SEP events is a complex interplay
 process with competition between the effects of particle scattering and adiabatic focusing. \label{evolution}}
\end{figure}

Figure \ref{evolution} presents the radial evolutions of the peak
intensities of 25 MeV (solid lines) and 32 MeV (dashed lines) proton
events, which are extracted from the simulation results in Figures
\ref{25MeV} and \ref{32MeV}, respectively. The solid and open
circles denote the peak intensities of the SEP events observed along
the magnetic field line (B-series) and along the radial direction
(A-series), respectively. The coverage of the source region of all
the SEP events is $45^{\circ}$ in longitude and latitude. We can
clearly see that the SEP peak intensities generally decrease with
increasing radial distances. However, the evolution process is
complicated and cannot be described by a single simple functional
form such as power-law function $R^{-\alpha}$ ($R$ is radial
distance). Specifically, for both particle energies 25 MeV and 32
MeV, the peak particle intensities of the B-series SEP cases
gradually decrease with increasing radial distances up to $\sim3.0$
AU, beyond where the peak intensities abruptly decrease to a very
low value, due to the significant reduction of the adiabatic
focusing effect at large radial distances \citep{He2012a}. For both
25 MeV and 32 MeV protons, the peak intensities of the A-series SEP
cases generally decrease with increasing radial distances up to
$\sim4.0$ AU, beyond where the peak SEP fluxes counterintuitively
increase with the increasing radial distances, due to the decreasing
longitudinal separations between the SEP source and the magnetic
footpoints of observers at $\gtrsim3.2$ AU. As we know, the farther
the magnetic footpoint of the observer is away from the SEP source
in longitude, the smaller the particle intensity (including peak
intensity) observed will be \citep{He2011}. In general, the radial
evolution process of SEP intensities is the manifestation of the
competitive combination of various fundamental mechanisms such as
particle scattering and adiabatic focusing. The competition between
scattering and adiabatic focusing is especially meaningful in the
outer heliosphere, where the interplay between these two effects may
lead to the counter-streaming particle beams \citep{He2015}.

\begin{figure}
 \epsscale{1.0}
 \plotone{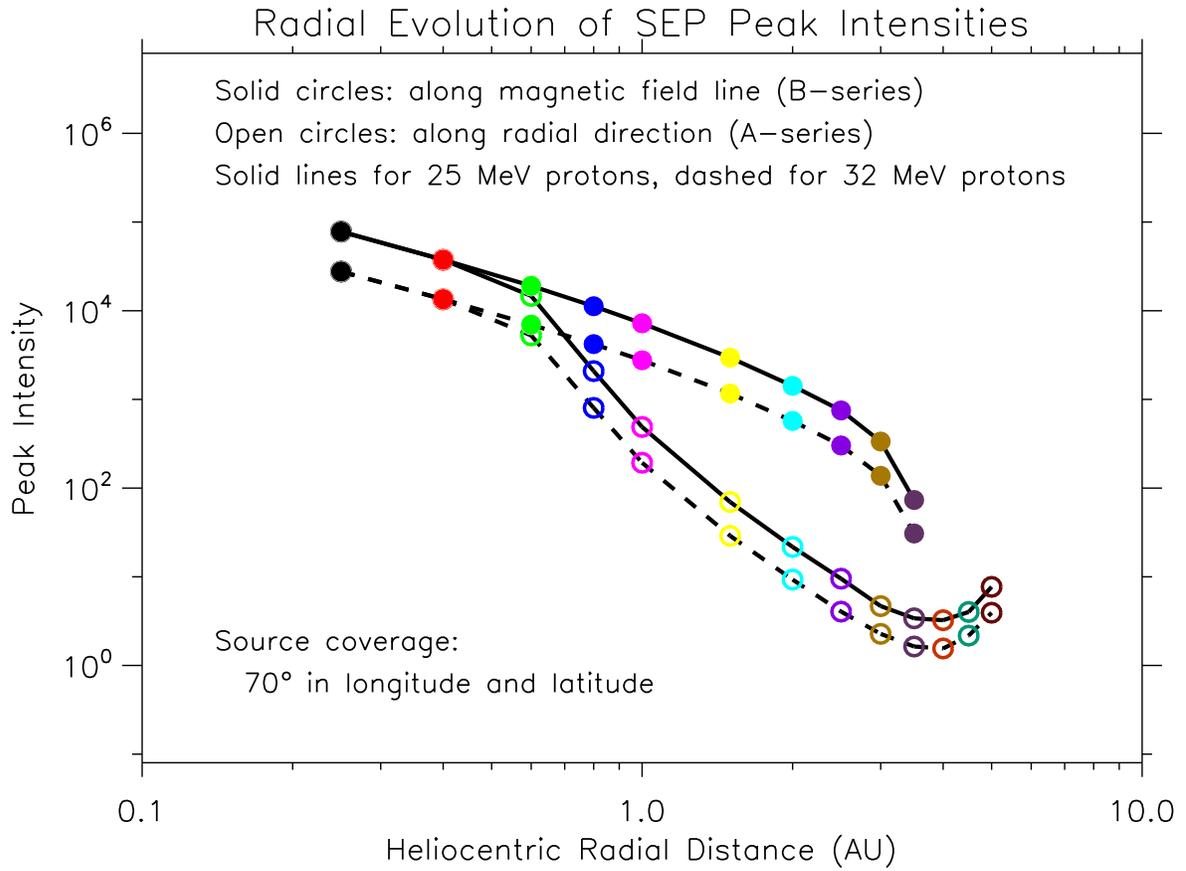}
 \caption{Same as Figure \ref{evolution} except for the SEP source of $70^{\circ}$ width in latitude and longitude. \label{evolution-big}}
\end{figure}

We also numerically simulate the SEP events with the source coverage
of $70^{\circ}$ in longitude and latitude. Figure
\ref{evolution-big} presents the simulation results of the radial
evolutions of such SEP cases. We note that except for the source
coverage, all of the other parameters of the simulations displayed
in Figure \ref{evolution-big} are the same as the parameters used in
Figure \ref{evolution}. As we can see, basically, the evolution
trend and evolution property of the SEP events presented in Figure
\ref{evolution-big} are similar to those in Figure \ref{evolution}.
In the beginning, the SEP peak intensities decrease with increasing
radial distances. However, the evolution process is complicated and
cannot be depicted by a single function such as $R^{-\alpha}$,
especially at large radial distances. For both energy channels, the
peak intensities in the B-series SEP cases decrease in a gradual
manner with increasing radial distances up to $\sim3.0$ AU, beyond
where the peak intensities suddenly decline to a quite low value,
due to the largely reduced focusing effect in the outer heliosphere.
In the A-series SEP cases, the peak intensities decrease with
increasing radial distances up to $\sim4.0$ AU, beyond where the
peak fluxes increase with the increasing radial distances, due to
the decreasing longitudinal separations between the particle source
and the magnetic footpoints of spacecraft at $\gtrsim3.2$ AU.
Because the decreasing longitudinal distances between SEP source and
spacecraft footpoints indicate increasing particle intensities
observed by these spacecraft \citep{He2011}. Therefore, the radial
evolution of SEP events is an interplay process between particle
scattering and adiabatic focusing.

In March 1987, a workshop on the interplanetary particle environment
was held at the Jet Propulsion Laboratory in Pasadena, California.
In this workshop, recommendations for radial extrapolation of peak
particle fluxes detected at 1 AU to other radial distances were
adopted by the working group consensus and read as follows
\citep{Feynman1988}:

1. To infer proton intensities at radial distances $R>1$ AU from the
intensity measurements at 1 AU, use a function $R^{-3.3}$ with
variations from $R^{-4}$ to $R^{-3}$.

2. To infer proton intensities at radial distances $R<1$ AU from the
intensity measurements at 1 AU, use a function $R^{-3}$ with
variations from $R^{-3}$ to $R^{-2}$.

From our simulation results, we can see that the radial evolution of
SEP events is a quite complicated process which incorporates several
fundamental mechanisms such as particle scattering and adiabatic
focusing. The evolution process cannot be simply described by a
functional form of $R^{-\alpha}$, especially in the outer
heliosphere, where the competitive interplay between the effects of
scattering and adiabatic focusing is quite considerable. Therefore,
the consensus recommendations for radial extrapolation of SEP
intensities empirically adopted during the 1987 workshop are
oversimple.

\section{Summary and Conclusion}
In this work, we investigate the three-dimensional propagation and
radial evolution of SEPs from the inner through outer heliosphere by
numerically solving the five-dimensional Fokker-Planck transport
equation incorporating the perpendicular diffusion mechanism. We
analyze the effects of adiabatic focusing and scattering on the SEP
intensities. The interplay process between these effects is
discussed in detail by investigating the radial evolution of SEP
time-intensity profiles. Some peculiar and interesting phenomena of
SEP transport in the three-dimensional interplanetary magnetic field
are found for the first time, to our knowledge. For instance, at
large radial distances from the Sun, the particle intensities in the
B-series SEP cases abruptly decline to a quite low value, and on the
contrary, the particle intensities in the A-series SEP cases
increase with the increasing radial distances. We discuss the
physical mechanisms responsible for the formation of these peculiar
SEP phenomena and conclude that these SEP evolution behaviors result
from the interplay process between particle scattering and adiabatic
focusing. We analyze the radial dependence of SEP peak intensities
from the inner through outer heliosphere and point out that it
cannot be described merely by a single functional form
$R^{-\alpha}$, especially at large radial distances. We also
numerically reproduce the famous SEP ``flood" (previously
``reservoir") phenomenon from the inner through outer heliosphere.
In addition, our findings can also be used to predict the
observations made by future missions in the interplanetary space.

%%%%%%%%%%%%%%%%%%%%%%%%%%%%%%%%%%%%%%%%%%%%%%%%%%%%%%%%%%%%%%%%%

\acknowledgments

This work was supported in part by the National Natural Science
Foundation of China under grants 41621063, 41874207, 41474154, and
41204130, and the Chinese Academy of Sciences under grant
KZZD-EW-01-2. H.-Q.H. gratefully acknowledges the partial support of
the Youth Innovation Promotion Association of the Chinese Academy of
Sciences (No. 2017091). We benefited from the energetic particle
data of IMP-8 and Ulysses provided by NASA/Space Physics Data
Facility (SPDF)/CDAWeb.

%%%%%%%%%%%%%%%%%%%%%%%%%%%%%%%%%%%%%%%%%%%%%%%%%%%%%%%%%%%%%%%%%

\clearpage

%%%%%%%%%%%%%%%%%%%%%%%%%%%%%%%%%%%%%%%%%%%%%%%%%%%%%%%%%%%%%%%%%

\clearpage

\clearpage

\clearpage

\clearpage

\clearpage

%%%%%%%%%%%%%%%%%%%%%%%%%%%%%%%%%%%%%%%%%%%%%%%%%%%%%%%%%%%%%%%%%

\end{document}